# Compliant Mechanism Synthesis Using Nonlinear Elastic Topology Optimization with Variable Boundary Conditions

Lee R. Alacoque[1*], Anurag Bhattacharyya[2], and Kai A. James[3]

In topology optimization of compliant mechanisms, the specific placement of boundary conditions strongly affects the resulting material distribution and performance of the design. At the same time, the most effective locations of the loads and supports are often difficult to find manually. This substantially limits topology optimization's effectiveness for many mechanism design problems. We remove this limitation by developing a method which automatically determines optimal positioning of a prescribed input displacement and a set of supports simultaneously with an optimal material layout. Using nonlinear elastic physics, we synthesize a variety of compliant mechanisms with large output displacements, snap-through responses, and prescribed output paths, producing designs with significantly improved performance in every case tested. Compared to optimal designs generated using best-guess boundary conditions used in previous studies, the mechanisms presented in this paper see performance increases ranging from 47%-380%. The results show that nonlinear mechanism responses may be particularly sensitive to boundary condition locations and that effective placements can be difficult to find without an automated method.

## 1. Introduction

Topology optimization [1] is a computational design method used to automatically find an optimal distribution of material that minimizes an objective function while satisfying a set of constraint functions. Originally introduced as a method for maximizing the stiffness of static structures [2], it required the user to specify the configuration of applied loads and rigid supports which would then remain fixed and unchanging during the optimization process. This feature persisted when topology optimization was later applied to the synthesis of compliant mechanisms [3], [4]. Compliant mechanism design has since become a common use of the method, with a wide variety of applications ranging from morphing wings [5] to micro-manipulators [6]–[9], flexures [10], switching mechanisms [11], [12], and more [13]. Yet it still remains standard practice for the boundary conditions to be predetermined by the user based only on intuition, despite the fact that in mechanism design the configuration of the boundary conditions has a significant influence on the resulting optimal geometry and its motion. Hence, topology optimization's effectiveness as a tool for designing compliant mechanisms can be significantly limited by requiring the user to manually choose boundary conditions.

Up until recently, efforts to incorporate the configuration of boundary conditions as automatically optimized parameters in topology optimization have been limited to only the supports, with the location and direction of the applied load still chosen by the user of the software. This has mostly been applied to static structures [14]–[18], with a few examples of compliant mechanism design [19]–[21] where gains of up to a 77% increase in mechanism performance were shown in the study by Buhl [19]. Optimizing the applied load location and direction in topology optimization to maximize the performance of structural designs was first accomplished by Alacoque and James [22], where the material distribution, support locations, load location, and load orientation were simultaneously optimized to maximize the output displacement of a compliant displacement inverter mechanism. A 150% gain in performance was achieved over the optimal design generated using the conventionally chosen boundary conditions. Since then, Lee and Xie have adopted similar methods for optimizing applied loading conditions [23], [24], achieving more efficient structural designs.

---

[1] University of Illinois Urbana-Champaign; Urbana, Illinois, USA

[2] Palo Alto Research Center; Palo Alto, California, USA

[3] Georgia Institute of Technology, Atlanta, Georgia, USA

[*] Corresponding author. Email: leea2@illinois.edu

The results for compliant mechanism design in [22] are promising, however the linear elastic modeling used in the study has certain limitations. It can be inaccurate when large rotations occur in the deformation of the mechanism, and it is only able to model straight-line motions and linear load-displacement curves. Topology optimization using nonlinear elastic physics accurately models large deformations and produces different optimal material distributions than with linear analysis [25]–[27]. It also makes it possible to synthesize compliant mechanisms with nonlinear behaviors such as bistability [28]–[30], tailored nonlinear load-displacement curves [31], [32], and nonlinear output paths [33], [34]. These behaviors are typically modeled using displacement-controlled algorithms, while the linear elastic studies [22]–[24] with variable boundary conditions apply the forces directly. To perform these interesting nonlinear optimizations with variable boundary conditions, a new method for a variable input displacement must be developed in addition to extending the previous methods to nonlinear elastic physics.

In this work we develop the formulations necessary for optimizing the boundary condition configuration simultaneously with material distribution in displacement-controlled, nonlinear elastic topology optimization. The super-Gaussian projection function method from [22] is extended to control the effective locations of the boundary conditions by scaling the magnitudes of body forces and elastic support spring constants applied to each finite element. For complete control over the input actuation behavior and to enable the optimization of different nonlinear behaviors, we formulate a new displacement-controlled iterative solver that enables a continuously variable input displacement location and orientation. Novel adjoint sensitivity equations are derived to accompany the solver, allowing the use of gradient-based optimizers such as the Method of Moving Asymptotes (MMA) [35]. We use our method to generate compliant mechanisms using several different types of objective function and compare the improved performance of the resulting designs to those obtained using best-guess boundary conditions from other nonlinear topology optimization studies in the literature.

The rest of the paper is organized as follows. Section 2 describes the density-based topology optimization formulation and the super-Gaussian projection method used to vary the boundary condition configuration. Section 3 describes the nonlinear, hyperelastic finite element formulation and the modifications to the residual and external load vector formulations that facilitate the variable-input displacement-control algorithm, which is outlined in Section 4. The adjoint sensitivity analysis is performed in Section 5 to derive generic derivative formulas that can be applied to any objective or constraint function. In Section 6 we present and analyze the optimization results, and in Section 7 we discuss conclusions and the possibilities for future research directions.

## 2. Topology Optimization with Boundary Conditions as Design Variables

The topology optimization problem we are considering is illustrated in Figure 1, which shows an elastic body in its undeformed configuration. The material density $\boldsymbol{\rho}$ is distributed within a design domain and is supported by fasteners at locations $\left(X_s^{(i)}, Y_s^{(i)}\right)$. A linear guided actuator applies a force to a location $\left(X_f, Y_f\right)$ to create a prescribed input displacement $\boldsymbol{U}_{in}$ at an angle $\theta$ and with stroke length $\|\boldsymbol{U}_{in}\|$. This input displacement causes an output displacement $U_{out}$ at some other point on the body. Circular fastening components at the load and support locations are represented by movable, solid, non-designable regions of material. Other fixed solid or void non-design regions may also be present in the domain. The goal is to find the material distribution, support locations, actuator location, and actuator angle which achieves the desired mechanism motion along with any specified structural characteristics. The design space is parameterized by a concatenated vector of the design variables, $\boldsymbol{\zeta}$, consisting of both the material density distribution and the boundary condition configuration:

$$\boldsymbol{\zeta} = [\boldsymbol{\rho} \quad \boldsymbol{X}_s \quad \boldsymbol{Y}_s \quad X_f \quad Y_f \quad \theta]. \tag{1}$$

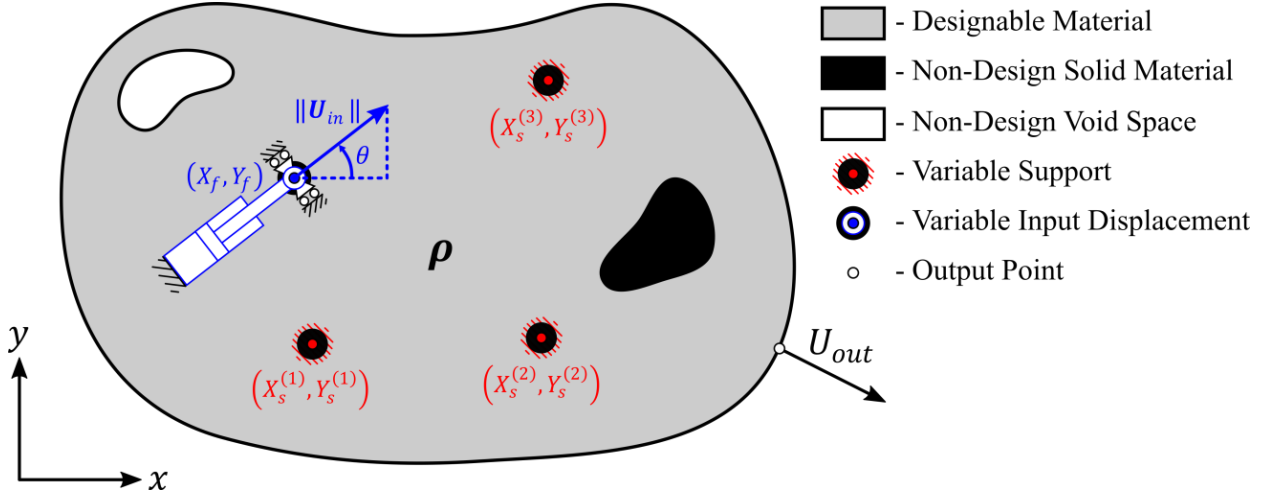

Figure 1 – The general design problem for compliant mechanisms with variable boundary conditions. The goal is to find a material distribution $\boldsymbol{\rho}$ along with support locations $\boldsymbol{X}_s, \boldsymbol{Y}_s$, actuator location $X_f, Y_f$, and actuator orientation $\theta$ which minimizes or maximizes a given objective function.

The continuous representation of the elasticity problem shown in Figure 1 is discretized using finite element analysis, creating a mesh of $N_e$ elements, shown in Figure 2. Each element $e$ in the mesh is assigned a material density design variable $\rho_e$ which can vary continuously from 0 to 1, representing an interpolation from empty space to solid material. We use the standard density filtering method of topology optimization to avoid checkerboard patterns and mesh dependence of designs [25]:

$$\tilde{\boldsymbol{\rho}} = \boldsymbol{W}\boldsymbol{\rho}, \tag{2}$$

where the vector $\tilde{\boldsymbol{\rho}}$ contains the filtered densities, obtained by multiplying the vector of density design variables with the $N_e \times N_e$ filter matrix $\boldsymbol{W}$. The entries of the filter matrix are calculated as

$$W_{ei} = \frac{w_{ei}}{\sum_{i=1}^{N_e} w_{ei}}, \tag{3}$$

where the weight factors $w_{ei}$ are computed based on the filter radius, $r_{min}$, and the distance between centroids of elements $e$ and $i$, $\Delta(e, i)$:

$$w_{ei} = \max\bigl(0, r_{min} - \Delta(e, i)\bigr). \tag{4}$$

Typically in topology optimization, the filtered densities are taken as the effective physical densities used to evaluate the element stiffness properties and solve the elasticity problem. However, in this work we perform one additional processing step to combine the filtered densities with the variable non-design regions at the boundary conditions points.

The boundary conditions must also be formulated as smooth and continuous functions of the design variables. To achieve this, elastic supports are applied everywhere in the domain with the intent that their stiffness will be varied to control where the structure is effectively restrained. Areas with supports of high stiffness will simulate rigid fixtures, and areas with supports of low or zero stiffness will leave the structure free to deform. To this end, linear elastic springs are connected from rigid boundaries to the elements' centroids to resist deformation in each degree of freedom. This adds no additional degrees of freedom to the finite element model. Each element is assigned a corresponding spring constant for its support, $k_e^s$, the value of which will depend on the support location design variables. With the same idea for the applied load, a force magnitude $f_e$ is assigned to each element centroid. The angle the forces are applied at will be determined by the displacement-control algorithm and is not the same as the input displacement angle $\theta$, since the actuator is guided and generates a lateral reaction force.

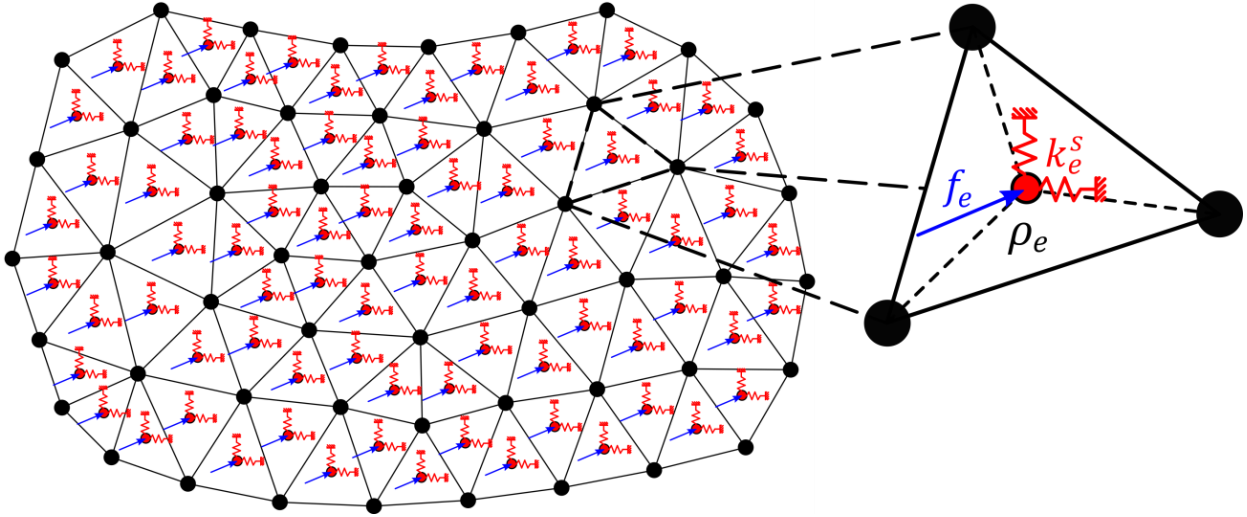

Figure 2 – A design domain discretized by finite elements. Forces are applied to each element centroid and spring elements connect the elements to rigid supports.

The effective locations of the boundary conditions are controlled using a super-Gaussian projection function [22], [36], which is a feature-mapping technique [37] based on distance functions. Given a field of distance to points or lines, the function outputs a field of a specified value within shapes of given radius that smoothly transition to zero in the direction of increasing distance. The function is plotted and labeled in Figure 3 for a simple one-dimensional distance function to a single point. Its general form is given by

$$G = Ab^{-\left(\frac{d^2}{r^2}\right)^P},$$ (5)

where $d$ is the distance field, $A$ is the value the function takes within each geometric feature, $P$ is a parameter that controls the sharpness of the transition region, and $r$ sets the length from zero distance to where the function equals the value specified by the parameter $b$:

$$G(d = r) = \frac{A}{b}.$$ (6)

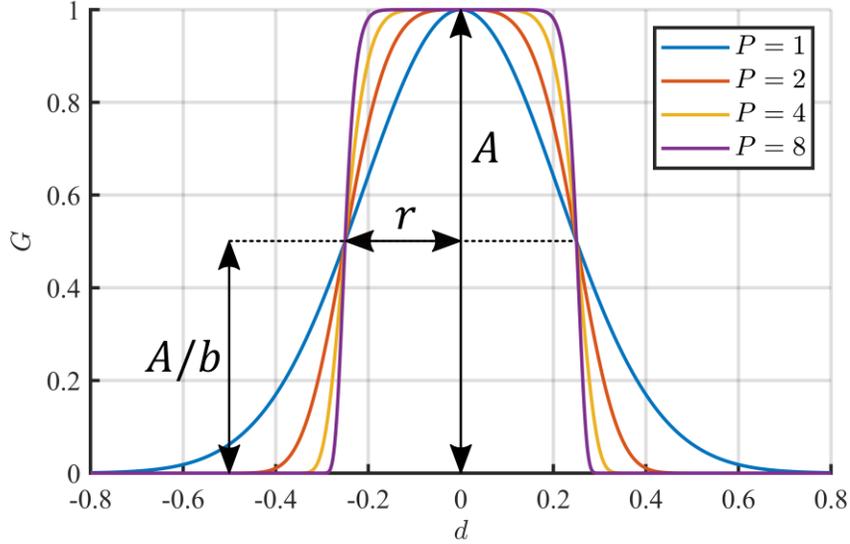

Figure 3 – The super-Gaussian projection function with a 1D distance function for $A = 1$, $b = 2$, $r = 0.25$, and several values of the exponent $P$.

For this work, we use the smoothed minimum distance from each element centroid $(\bar{X}_e, \bar{Y}_e)$ to a number of points $(X_i, Y_i)$ placed within the design domain:

$$d_e = \left( \sum_i \left( \sqrt{(\bar{X}_e - X_i)^2 + (\bar{Y}_e - Y_i)^2} \right)^{-Q} \right)^{-\frac{1}{Q}},$$

(7)

where larger values of $Q$ more closely approximate the true minimum. An example of this distance field, and the circular features the super-Gaussian projection function creates from it, is shown in Figure 4.

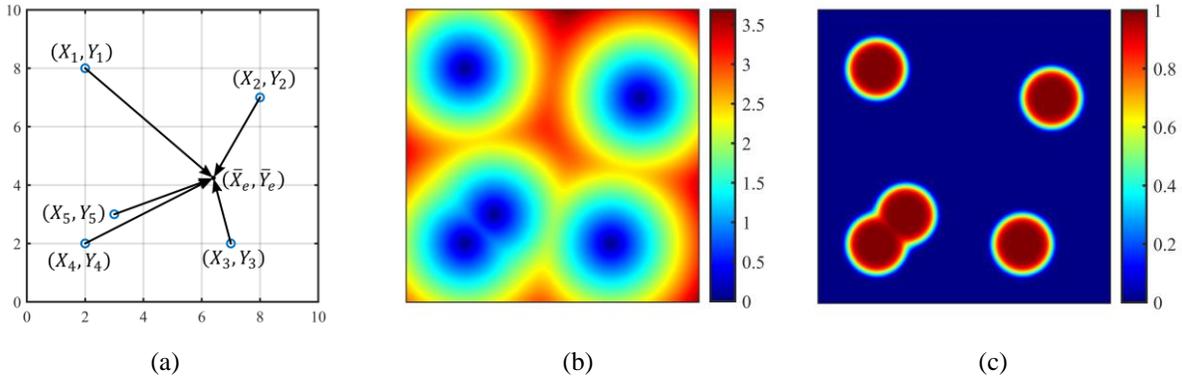

| (a) | (b) | (c) |

Figure 4 – Projecting circular shapes from zero dimensional points. (a) Several points in the domain with their spatial coordinates and distances to the centroid of an element $e$. (b) The smooth distance field for the set of points, using $Q = 10$. (c) The super-Gaussian projection of the smooth distance field, using $A = 1$, $b = 2$, $r = 1$, and $P = 4$.

Now applying the projection function to control the distribution of elastic support stiffness, the design variables $\boldsymbol{X}_s$ and $\boldsymbol{Y}_s$ are used in Equation (7) to create the smoothed distance field. The coefficient $A$ from Equation (5) is chosen such that the spring constants within the support geometry are equivalent to the shear stiffness of a solid support material with shear modulus $G_s$ and thickness $t_s$. Considering an element of area $A_e$ with a force $f_e$ applied to the centroid of its cross section, the average shear stress in its support material is

$$\tau_e = \frac{f_e}{A_e}. \tag{8}$$

Using a linear stress-strain relationship and assuming a small transverse displacement $\delta_e$ gives:

$$\frac{\delta_e}{t_s} G_s = \frac{f_e}{A_e}. \tag{9}$$

The equivalent spring constant for the solid support material of the element is determined by rearranging and comparing to the equation for a spring:

$$\frac{G_s}{t_s} A_e \delta_e = f_e. \tag{10}$$

Substituting the spring constant for the coefficient of the super-Gaussian function yields:

$$k_e^s = \frac{G_s}{t_s} A_e b^{-\left(\frac{d_e^2}{r^2}\right)^P}. \tag{11}$$

For the applied load, the design variables $X_f$ and $Y_f$ are used for the distance function:

$$d_e = \sqrt{\left(\overline{X}_e - X_f\right)^2 + \left(\overline{Y}_e - Y_f\right)^2}. \tag{12}$$

The distribution of load magnitude is

$$f_e = A_f b^{-\left(\frac{d_e^2}{r^2}\right)^P}, \tag{13}$$

where the coefficient $A_f$ is chosen such that the total load applied in the design domain is equal to 1:

$$A_f = \frac{1}{\sum_{e=1}^{N_e} b^{-\left(\frac{d_e^2}{r^2}\right)^P} V_e}. \tag{14}$$

This is assumed to be a constant and is only calculated once before beginning the optimization.

Moveable non-design regions around the boundary condition points are used to ensure well-defined compliant hinges are generated at the supports [22], and so that the forces from the actuator are always applied to solid material. To create them, a density distribution is projected onto the mesh using the super-Gaussian function. The distance function is created using both the support location and load location design variables. The coefficient $A$ is set to 1 for solid material, giving the projected density distribution as

$$\hat{\rho}_e = b^{-\left(\frac{d_e^2}{r^2}\right)^P}. \tag{15}$$

The projected densities and the filtered densities are then combined into the physical density field using a smooth maximum function:

$$\bar{\rho}_e = \left(\tilde{\rho}_e^Q + \hat{\rho}_e^Q\right)^{\frac{1}{Q}}.$$ (16)

The physical densities in each element, $\bar{\rho}_e$, define the design that is analyzed in the finite element analysis and is the density field that is presented as the results of topology optimization. It is used to calculate the elastic modulus of each element using the Solid Isotropic Material with Penalization (SIMP) interpolation scheme:

$$E_e = E_{min} + \bar{\rho}_e^p (E_0 - E_{min}),$$ (17)

where $E_{min}$ is a small value of stiffness given to void elements to avoid singular stiffness matrices in the finite element analysis, $E_0$ is the elastic modulus of the solid material, and $p$ is the SIMP penalization factor which reduces the stiffness-to-weight ratio of intermediate density elements causing the optimizations to converge to structures made up of mostly solid material.

## 3. Nonlinear Finite Element Model

In each iteration of the topology optimization process, nonlinear finite element analysis is used to simulate the deformation of the current design. The results of the analysis are then used to characterize the performance of the mechanism by calculating the values of the objective and constraint functions. The general theory and background of nonlinear finite element analysis can be found in references such as [38]–[40], and here we only present the equations that are needed for implementation. We use the total Lagrangian formulation and a hyperelastic material model to represent compliant mechanisms made from a rubber-like flexible material.

Typically, displacement-controlled nonlinear finite element analysis finds the vector of nodal displacements, $\boldsymbol{U}$, and a load intensity factor, $\lambda$, describing an equilibrium configuration of the structure where the externally applied loads are balanced with the internally generated forces. The balance of forces is represented by the residual vector:

$$\boldsymbol{R} = \lambda \boldsymbol{F}^{ext} - \boldsymbol{F}^{int} = \boldsymbol{0},$$ (18)

where $\boldsymbol{F}^{ext}$ is a reference vector of external nodal loads and $\boldsymbol{F}^{int}$ is the vector of internal nodal forces. For the work here, considering the applied element forces $f_e$ to be body forces per unit volume, the reference external load vector is given as an assembly of the body forces [41]:

$$\boldsymbol{F}^{ext} = \bigwedge_{e=1}^{N_e} \int_{V_e} \boldsymbol{N}^T f_e \begin{bmatrix} \cos(\phi) \\ \sin(\phi) \end{bmatrix} dV_e,$$ (19)

where the symbol $\bigwedge$ denotes the finite element assembly operation, $V_e$ is the volume of the element, the matrix $\boldsymbol{N}$ contains the finite element shape functions, and $\phi$ is the (unknown) angle the load is applied at. Assuming that the element force $f_e$ is evenly distributed to the nodes results in:

$$\boldsymbol{F}^{ext} = \bigwedge_{e=1}^{N_e} \frac{f_e}{n_n} \boldsymbol{\Phi} V_e,$$ (20)

where $n_n$ is the number of nodes in the element and $\mathbf{\Phi} = [\cos(\phi) \quad \sin(\phi) \quad ...]^T$ is a direction vector that applies the force components to the appropriate degrees of freedom. Since the load angle $\phi$ that causes the given input displacement angle is unknown, it must be found by the displacement-control algorithm. To facilitate this, the external load vector is split into $x$ and $y$ components and the residual vector is rewritten as

$$\boldsymbol{R} = \lambda_x \boldsymbol{F}_x^{ext} + \lambda_y \boldsymbol{F}_y^{ext} - \boldsymbol{F}^{int} = \boldsymbol{0},\tag{21}$$

where

$$\boldsymbol{F}_x^{ext} = \bigwedge_{e=1}^{N_e} \frac{f_e}{\sqrt{2}n_n} \mathbf{\Phi}_x V_e,\tag{22}$$

$$\boldsymbol{F}_y^{ext} = \bigwedge_{e=1}^{N_e} \frac{f_e}{\sqrt{2}n_n} \mathbf{\Phi}_y V_e,\tag{23}$$

with the direction vectors as $\mathbf{\Phi}_x = [1 \quad 0 \quad ...]^T$ and $\mathbf{\Phi}_y = [0 \quad 1 \quad ...]^T$. The two load intensity factors $\lambda_x$ and $\lambda_y$ are then determined by the displacement-control algorithm such that the specified input displacement $\boldsymbol{U}_{in}$ is achieved. This algorithm is described in the next section.

In topology optimization with nonlinear finite element analysis, a numerical issue occurs where elements with low stiffness can distort excessively and prevent convergence of the iterative Newton-Raphson solver. Earlier works circumvented this issue by ignoring nodes surrounded by void elements in the solver's convergence criteria [26], [33], by removing elements with low densities from the mesh [42], or by optimizing the connectivity of completely solid elements [43]. More recent methods include interpolating to linear elastic modeling at low densities [44], adding an extra hyperelastic stiffness term to unstable elements [45], [46], and stabilizing low-density elements in a way that is also able to simulate contact between solid regions [47]. Here, we use the linear-nonlinear strain energy interpolation scheme by Wang et al. [44] since it is relatively simple to implement and is able to converge for adequately large deformations in most cases. The interpolation function in this method is a smoothed Heaviside step function that is dependent on the physical density of each element:

$$\gamma_e = \frac{\tanh(\beta\rho_0) + \tanh(\bar{\rho}_e{}^p - \rho_0)}{\tanh(\beta\rho_0) + \tanh(\beta(1-\rho_0))},\tag{24}$$

where the parameter $\beta$ affects the sharpness of the step function and $\rho_0$ controls the location of the threshold. The deformation gradient in each element is computed as:

$$\boldsymbol{F} = \boldsymbol{I} + \gamma_e \nabla \boldsymbol{U}_e,\tag{25}$$

where $\boldsymbol{I}$ is the identity matrix. Like in the original implementation of the method [44], we use a modified Neo-Hookean hyperelastic material model [48] where the second Piola-Kirchhoff stress is implemented as:

$$\boldsymbol{S} = \lambda_0(2J^2 - J)\boldsymbol{C}^{-1} + \mu_0(\boldsymbol{I} - \boldsymbol{C}^{-1}),\tag{26}$$

and the entries of the constitutive matrix as:

$$D_{ijkl} = \lambda_0(2J^2 - J)C_{ij}^{-1}C_{kl}^{-1} + (\mu_0 - \lambda_0(2J^2 - J))(C_{ik}^{-1}C_{jl}^{-1} + C_{il}^{-1}C_{jk}^{-1}).\tag{27}$$

The constants $\lambda_0$ and $\mu_0$ are the Lamé parameters for a unit elastic modulus, $J$ is the determinant of the deformation gradient, and $\boldsymbol{C} = \boldsymbol{F}^T \boldsymbol{F}$ is the right Cauchy-Green deformation tensor. This interpolation method allows the Newton-Raphson iterations to converge in many cases, however intermediate density elements may still become unstable if displacements are too large. Low-stiffness elements may also become unstable if external loads are applied to them, which is prevented by the moveable solid non-design region placed under the input point.

The internal force vector is an assembly of the element force vectors for the continuum part of the mesh, interpolated between nonlinear and linear modeling, plus the internal forces of the support springs. Writing the nonlinear element force vectors as $\boldsymbol{f}_e^{NL}$ and the linear counterparts as $\boldsymbol{f}_e^L$, the global internal force vector is assembled as:

$$\boldsymbol{F}^{int} = \bigwedge_{e=1}^{N_e} (\gamma_e \boldsymbol{f}_e^{NL} + (1 - \gamma_e^2) \boldsymbol{f}_e^L) + \boldsymbol{K}_s \boldsymbol{U}, \tag{28}$$

where $\boldsymbol{K}_s$ is the global stiffness matrix of the supports, assembled as:

$$\boldsymbol{K}_s = \bigwedge_{e=1}^{N_e} \frac{k_e^S}{n_n} \boldsymbol{I}.$$

The size of the identity matrix $\boldsymbol{I}$ is equal to the number of degrees of freedom in the element. The nonlinear element force vector is given by the integration

$$\boldsymbol{f}_e^{NL} = E_e \int_{V_e} \boldsymbol{B}_N^T \boldsymbol{S} \, dV_e, \tag{29}$$

where $\boldsymbol{B}_N$ is the nonlinear strain-displacement matrix. The linear element force vector is given by

$$\boldsymbol{f}_e^L = E_e \int_{V_e} \boldsymbol{B}^T \boldsymbol{D}_0 \boldsymbol{\varepsilon} \, dV_e, \tag{30}$$

where $\boldsymbol{B}$ is the linear strain-displacement matrix, $\boldsymbol{D}_0$ is the linear constitutive matrix for unit stiffness, and $\boldsymbol{\varepsilon}$ is the linear strain in the element.

The tangent stiffness matrix is also an interpolation between nonlinear and linear modeling for the continuum part of the mesh, with the contribution from the linear elastic supports added to it:

$$\boldsymbol{K}_T = \bigwedge_{e=1}^{N_e} (\gamma_e^2 \boldsymbol{k}_e^{NL} + (1 - \gamma_e^2) \boldsymbol{k}_e^L) + \boldsymbol{K}_s. \tag{31}$$

The nonlinear element stiffness matrix is

$$\boldsymbol{k}_e^{NL} = E_e \int_{V_e} (\boldsymbol{B}_N^T \boldsymbol{D} \boldsymbol{B}_N + \boldsymbol{B}_G^T \boldsymbol{\Sigma} \boldsymbol{B}_G) \, dV_e, \tag{32}$$

where $\boldsymbol{\Sigma}$ is a matrix of second Piola-Kirchhoff stresses and $\boldsymbol{B}_G$ is another strain-displacement matrix [38]. The linear element stiffness matrix is integrated using

$$\boldsymbol{k}_e^L = E_e \int_{V_e} \boldsymbol{B}^T \boldsymbol{D_0} \boldsymbol{B} \, dV_e. \tag{33}$$

## 4. The Displacement-Control Algorithm

The displacement-control algorithm is an incremental-iterative (predictor-corrector) method [49], [50], where the input displacement is incremented in steps to obtain points along the load-displacement curve. For each increment, an iterative cycle solves for the unknown state variables $\boldsymbol{U}$, $\lambda_x$, and $\lambda_y$ to restore static equilibrium to within a given tolerance. The residual (Equation (21)) provides the constraint functions necessary to solve for the unknown nodal displacements, $\boldsymbol{U}$. However, an additional two are needed to solve for the load intensity factors, $\lambda_x$ and $\lambda_y$. Since the input displacement location is continuously variable and can be applied anywhere within the mesh, including at the interior of elements, it provides another two data points in the displacement field that can be used to define the extra constraint functions. The displacement components at the input actuation point are obtained by using the finite element shape functions:

$$\begin{bmatrix} U_x^{in}(X_f, Y_f) \\ U_y^{in}(X_f, Y_f) \end{bmatrix} = \boldsymbol{N}(X_f, Y_f)\boldsymbol{U}. \tag{34}$$

Then, the constraints can be defined as the difference between the displacement field at the input actuation point and the given input displacement, the two of which will be equal to each other at equilibrium:

$$\boldsymbol{c} = \boldsymbol{N}(X_f, Y_f)\boldsymbol{U} - \boldsymbol{U}_{in} = \begin{bmatrix} U_x^{in}(X_f, Y_f) - \|\boldsymbol{U}_{in}\|\cos(\theta) \\ U_y^{in}(X_f, Y_f) - \|\boldsymbol{U}_{in}\|\sin(\theta) \end{bmatrix} = \begin{bmatrix} 0 \\ 0 \end{bmatrix} \tag{35}$$

The displacement-control algorithm proceeds as follows. For each displacement step $i$, the predictor phase begins by calculating reference displacement increment vectors from the reference load vectors:

$$[\boldsymbol{K}_T]^{i-1}[\Delta \boldsymbol{U}^a \quad \Delta \boldsymbol{U}^b] = [\boldsymbol{F}_x^{ext} \quad \boldsymbol{F}_y^{ext}], \tag{36}$$

where the state at increment $i-1$ is from the previously converged displacement step, or else is the unloaded and undeformed state of the structure. With a given displacement step size $\Delta \overline{U}$, which is some fraction of the final input displacement length $\|\boldsymbol{U}_{in}\|$, the predictor load intensity increments are found by solving the following system of equations:

$$\begin{bmatrix} \Delta U_x^a(X_f, Y_f) & \Delta U_x^b(X_f, Y_f) \\ \Delta U_y^a(X_f, Y_f) & \Delta U_y^b(X_f, Y_f) \end{bmatrix} \begin{bmatrix} \Delta \lambda_x \\ \Delta \lambda_y \end{bmatrix} = \begin{bmatrix} \Delta \overline{U} \cos(\theta) \\ \Delta \overline{U} \sin(\theta) \end{bmatrix}, \tag{37}$$

where the coefficient matrix contains the $x$ and $y$ components of the displacement increment vectors taken at the input actuation location. These are found by using the finite element shape functions, similar to Equation (34). The last step in the predictor phase is to update the state variables as:

$$\boldsymbol{U}^i = \boldsymbol{U}^{i-1} + \Delta \lambda_x \Delta \boldsymbol{U}^a + \Delta \lambda_y \Delta \boldsymbol{U}^b, \tag{38}$$

$$\lambda_x^i = \lambda_x^{i-1} + \Delta \lambda_x, \tag{39}$$

$$\lambda_y^i = \lambda_y^{i-1} + \Delta \lambda_y. \tag{40}$$

The tangent stiffness matrix, internal force vector, and residual vector are also updated at this point using the state of the structure predicted in Equations (38)-(40).

If the norm of the residual is greater than a given tolerance, the algorithm moves into an inner loop. This is the corrector phase. For corrector iteration $j$, three reference displacement increment vectors are calculated using:

$$[\boldsymbol{K}_T]^j [\Delta \boldsymbol{U}^a \quad \Delta \boldsymbol{U}^b \quad \Delta \boldsymbol{U}^c]^j = \begin{bmatrix} \boldsymbol{F}_x^{ext} & \boldsymbol{F}_y^{ext} & \boldsymbol{R}^j \end{bmatrix}. \tag{41}$$

Using the shape functions to get the values of the reference displacement increment vectors at the input load location, the load intensity increments are solved for using:

$$\begin{bmatrix} \Delta U_x^a(X_f, Y_f) & \Delta U_x^b(X_f, Y_f) \\ \Delta U_y^a(X_f, Y_f) & \Delta U_y^b(X_f, Y_f) \end{bmatrix}^j \begin{bmatrix} \Delta \lambda_x \\ \Delta \lambda_y \end{bmatrix}^j = \begin{bmatrix} -\Delta U_x^c(X_f, Y_f) \\ -\Delta U_y^c(X_f, Y_f) \end{bmatrix}^j. \tag{42}$$

The total displacement increment vector is then given by

$$\Delta \boldsymbol{U}^j = \Delta \lambda_x^j \Delta \boldsymbol{U}_j^a + \Delta \lambda_y^j \Delta \boldsymbol{U}_j^b + \Delta \boldsymbol{U}_j^c, \tag{43}$$

and the state variables are updated as:

$$\boldsymbol{U}^{i,j+1} = \boldsymbol{U}^{i,j} + \Delta \boldsymbol{U}^j, \tag{44}$$

$$\lambda_x^{i,j+1} = \lambda_x^{i,j} + \Delta \lambda_x^j, \tag{45}$$

$$\lambda_y^{i,j+1} = \lambda_y^{i,j} + \Delta \lambda_y^j. \tag{46}$$

The tangent stiffness matrix and internal force vector are reassembled again here, and the residual vector is recalculated. The corrector phase continues to iterate until the residual norm convergence criterion is satisfied, meaning the displacement step is converged, and the algorithm then moves to increment $i + 1$. Once the algorithm converges at the given final displacement $\boldsymbol{U}_{in}$, it ends and outputs the results.

The corrector can fail to converge for various reasons. Possible causes include that the chosen step size is too large, that deformations larger than the interpolation scheme [44] is able handle have caused intermediate density elements to become unstable, or that a displacement limit point in the load-displacement curve has been encountered. To avoid these issues as much as possible, the algorithm is written to automatically try again with a bisected step length if the corrector passes a given maximum number of iterations. Resolving these issues would require better methods for stabilizing low-density elements, and a different solver such as an arc-length method [51] that can trace the load-displacement curve through limit points. These problems are left for future research.

## 5. Adjoint Sensitivity Analysis

Gradient-based methods of optimization require the derivatives of the objective and constraint functions with respect to each design variable. In topology optimization, the large number of design variables and the computational expense of finite element analysis means that the gradient computations must be performed efficiently. The adjoint sensitivity analysis method is used here to derive analytical sensitivity formulas that are inexpensive to evaluate.

For any objective or constraint function $f(\boldsymbol{\zeta}, \boldsymbol{\eta}(\boldsymbol{\zeta}))$, which is considered to be an *explicit* function of the design variables $\boldsymbol{\zeta}$ and the state variables $\boldsymbol{\eta}$, and where the state variables are considered *implicit* functions of the design variables, an augmented Lagrangian function $g(\boldsymbol{\zeta}, \boldsymbol{\eta}(\boldsymbol{\zeta}))$ is formed by multiplying the governing equations by Lagrange multipliers and adding these terms to the function $f$. Since the governing equations are equal to zero at equilibrium, the augmented Lagrangian function $g$ is equivalent to the original function $f$.

In conventional displacement-controlled topology optimization with constant boundary conditions applied to discrete nodes, the residual vector contains all of the governing equations. These include the equations for the *free* degrees of freedom, where the displacements are unknown and are solved for, and the equations for the *prescribed* degrees of freedom, where the displacements are given while the external loads are the unknowns [25]. However, in the context of variable boundary conditions (the subject of this work), all degrees of freedom in the mesh are "free" with unknown displacements. Thus, only adding the residual vector in the augmented Lagrangian function is not enough to capture the total derivative, and another term is needed. This missing term of governing equations is the displacement-control solver's constraint functions, $\boldsymbol{c}$, that were defined in Equation (35).

With the state variables $\boldsymbol{\eta}(\boldsymbol{\zeta}) = [\boldsymbol{U}(\boldsymbol{\zeta}) \quad \boldsymbol{\lambda}(\boldsymbol{\zeta})]^T$, where $\boldsymbol{\lambda}(\boldsymbol{\zeta}) = [\lambda_x(\boldsymbol{\zeta}) \quad \lambda_y(\boldsymbol{\zeta})]^T$, the augmented Lagrangian function is formed by multiplying the residual and the solver constraint functions with Lagrange multipliers $\boldsymbol{\psi}_R$ and $\boldsymbol{\psi}_c$, respectively, and adding the terms to the original function:

$$g\big(\boldsymbol{\zeta}, \boldsymbol{U}(\boldsymbol{\zeta}), \boldsymbol{\lambda}(\boldsymbol{\zeta})\big) = f\big(\boldsymbol{\zeta}, \boldsymbol{U}(\boldsymbol{\zeta}), \boldsymbol{\lambda}(\boldsymbol{\zeta})\big) + \boldsymbol{\psi}_R^T \boldsymbol{R}\big(\boldsymbol{\zeta}, \boldsymbol{U}(\boldsymbol{\zeta}), \boldsymbol{\lambda}(\boldsymbol{\zeta})\big) + \boldsymbol{\psi}_c^T \boldsymbol{c}\big(\boldsymbol{\zeta}, \boldsymbol{U}(\boldsymbol{\zeta}), \boldsymbol{\lambda}(\boldsymbol{\zeta})\big). \tag{47}$$

Denoting total or implicit derivatives by the operator $d/d\zeta$, and partial or explicit derivatives by the operator $\partial/\partial\zeta$, the total derivative is taken with respect to an arbitrary design variable using the multivariable chain rule of calculus:

$$\frac{dg}{d\zeta} = \frac{\partial f}{\partial \zeta} + \frac{\partial f}{\partial \boldsymbol{U}}\frac{d\boldsymbol{U}}{d\zeta} + \frac{\partial f}{\partial \boldsymbol{\lambda}}\frac{d\boldsymbol{\lambda}}{d\zeta} + \boldsymbol{\psi}_R^T\left(\frac{\partial \boldsymbol{R}}{\partial \zeta} + \frac{\partial \boldsymbol{R}}{\partial \boldsymbol{U}}\frac{d\boldsymbol{U}}{d\zeta} + \frac{\partial \boldsymbol{R}}{\partial \boldsymbol{\lambda}}\frac{d\boldsymbol{\lambda}}{d\zeta}\right) + \boldsymbol{\psi}_c^T\left(\frac{\partial \boldsymbol{c}}{\partial \zeta} + \frac{\partial \boldsymbol{c}}{\partial \boldsymbol{U}}\frac{d\boldsymbol{U}}{d\zeta} + \frac{\partial \boldsymbol{c}}{\partial \boldsymbol{\lambda}}\frac{d\boldsymbol{\lambda}}{d\zeta}\right). \tag{48}$$

This expression is rearranged to isolate the implicit derivatives of the state variables:

$$\frac{dg}{d\zeta} = \frac{\partial f}{\partial \zeta} + \boldsymbol{\psi}_R^T\frac{\partial \boldsymbol{R}}{\partial \zeta} + \boldsymbol{\psi}_c^T\frac{\partial \boldsymbol{c}}{\partial \zeta} + \left(\frac{\partial f}{\partial \boldsymbol{U}} + \boldsymbol{\psi}_R^T\frac{\partial \boldsymbol{R}}{\partial \boldsymbol{U}} + \boldsymbol{\psi}_c^T\frac{\partial \boldsymbol{c}}{\partial \boldsymbol{U}}\right)\frac{d\boldsymbol{U}}{d\zeta} + \left(\frac{\partial f}{\partial \boldsymbol{\lambda}} + \boldsymbol{\psi}_R^T\frac{\partial \boldsymbol{R}}{\partial \boldsymbol{\lambda}} + \boldsymbol{\psi}_c^T\frac{\partial \boldsymbol{c}}{\partial \boldsymbol{\lambda}}\right)\frac{d\boldsymbol{\lambda}}{d\zeta}. \tag{49}$$

Since the residual vector and solver constraint functions are equal to zero, the values of the Lagrange multipliers are arbitrary and can be chosen freely without affecting the value of the sensitivity. The terms with the implicit derivatives can be eliminated by finding Lagrange multipliers that cause the sum of the terms inside the brackets to vanish:

$$\frac{\partial f}{\partial \boldsymbol{U}} + \boldsymbol{\psi}_R^T\frac{\partial \boldsymbol{R}}{\partial \boldsymbol{U}} + \boldsymbol{\psi}_c^T\frac{\partial \boldsymbol{c}}{\partial \boldsymbol{U}} = \boldsymbol{0}, \tag{50}$$

$$\frac{\partial f}{\partial \boldsymbol{\lambda}} + \boldsymbol{\psi}_R^T\frac{\partial \boldsymbol{R}}{\partial \boldsymbol{\lambda}} + \boldsymbol{\psi}_c^T\frac{\partial \boldsymbol{c}}{\partial \boldsymbol{\lambda}} = \boldsymbol{0}. \tag{51}$$

Solving Equations (50) and (51) simultaneously leads to the following systems of linear equations that yield the values of the Lagrange multipliers:

$$\boldsymbol{\psi}_c^T\left(\frac{\partial \boldsymbol{c}}{\partial \boldsymbol{\lambda}} - \left(\frac{\partial \boldsymbol{c}}{\partial \boldsymbol{U}}\frac{\partial \boldsymbol{R}^{-1}}{\partial \boldsymbol{U}}\right)\frac{\partial \boldsymbol{R}}{\partial \boldsymbol{\lambda}}\right) = -\left(\frac{\partial f}{\partial \boldsymbol{\lambda}} - \left(\frac{\partial f}{\partial \boldsymbol{U}}\frac{\partial \boldsymbol{R}^{-1}}{\partial \boldsymbol{U}}\right)\frac{\partial \boldsymbol{R}}{\partial \boldsymbol{\lambda}}\right), \tag{52}$$

$$\boldsymbol{\psi}_R^T\frac{\partial \boldsymbol{R}}{\partial \boldsymbol{U}} = -\left(\frac{\partial f}{\partial \boldsymbol{U}} + \boldsymbol{\psi}_c^T\frac{\partial \boldsymbol{c}}{\partial \boldsymbol{U}}\right), \tag{53}$$

where

$$\frac{\partial \boldsymbol{R}}{\partial \boldsymbol{U}} = -\boldsymbol{K}_T, \tag{54}$$

$$\frac{\partial \boldsymbol{R}}{\partial \boldsymbol{\lambda}} = \begin{bmatrix} \boldsymbol{F}_x^{ext} & \boldsymbol{F}_y^{ext} \end{bmatrix}, \tag{55}$$

$$\frac{\partial \mathbf{c}}{\partial \boldsymbol{\lambda}} = \boldsymbol{0}, \tag{56}$$

$$\frac{\partial \mathbf{c}}{\partial \boldsymbol{U}} = \boldsymbol{N}(X_f, Y_f). \tag{57}$$

Now, without ever having had to derive expressions for the implicit derivatives of the iteratively computed state variables, the total derivative of any function $f$ is given by Equations (52)-(57) and the following formula:

$$\frac{dg}{d\zeta} = \frac{\partial f}{\partial \zeta} + \boldsymbol{\psi}_R^T \frac{\partial \boldsymbol{R}}{\partial \zeta} + \boldsymbol{\psi}_c^T \frac{\partial \mathbf{c}}{\partial \zeta}. \tag{58}$$

Depending on the specific function $f$ being implemented, only the three explicit derivatives $\partial f / \partial \zeta$, $\partial f / \partial \boldsymbol{U}$, and $\partial f / \partial \boldsymbol{\lambda}$ need to be found and plugged into Equations (52), (53), and (58), which is typically simple to do.

We note that the derivatives of the shape functions with respect to the input displacement coordinates, which appear in the derivative $\partial \mathbf{c} / \partial \zeta$, are discontinuous for shape functions that are not smooth across element boundaries. However, for our implementation using first-order elements with piecewise linear shape functions, we did not encounter any oscillations or other problems in the convergence of the topology optimization process that could be attributable to this.

## 6. Numerical Examples

In this section, the framework described in Sections 2 through 5 is implemented in MATLAB and is used to generate several types of compliant mechanisms based on examples from nonlinear topology optimization in the literature. Three types of design problems are investigated: displacement maximization, bistability, and path generation. We run each example twice: once with the boundary conditions fixed at the initial locations chosen by the other studies, and the second time allowing our algorithm to automatically adjust them to find optimal placements. The starting distribution of material is uniform in both cases. The performance of the designs obtained using variable boundary conditions is compared to those obtained using fixed boundary conditions to quantify the advantages of the method over conventional topology optimization.

Several quantities of interest are used to define the various objective and constraint functions in the example problems. The displacements at the output points of the mechanisms, at any point $m$ on the load-displacement curve where the state variables have been solved for, are found using

$$U_{out}^{(m)} = \boldsymbol{L}^T \boldsymbol{U}^{(m)}, \tag{59}$$

where $\boldsymbol{L}$ is a vector of all zeros, except for a value of one at the degrees of freedom that are selected by the user. The input force applied by the actuator is in-line with the input displacement and is found from the applied force's $x$ and $y$ components as

$$F_{in}^{(m)} = \sqrt{\left(\lambda_x^{(m)}\right)^2 + \left(\lambda_y^{(m)}\right)^2} \sin\left(\theta + \tan^{-1}\left(\frac{\lambda_x^{(m)}}{\lambda_y^{(m)}}\right)\right), \tag{60}$$

and the reaction force generated by the guide structure is the force perpendicular to the input displacement, given by

$$F_p^{(m)} = \sqrt{\left(\lambda_x^{(m)}\right)^2 + \left(\lambda_y^{(m)}\right)^2} \cos\left(\theta + \tan^{-1}\left(\frac{\lambda_x^{(m)}}{\lambda_y^{(m)}}\right)\right). \tag{61}$$

The volume fraction of material distributed in the domain is found from the physical densities:

$$V_f = \frac{\sum_{e=1}^{N_e} \bar{\rho}_e V_e}{\sum_{e=1}^{N_e} V_e}. \tag{62}$$

These quantities can be used in many different ways to control the characteristics of the mechanisms that are generated. The output displacements can be used in objective functions to create large geometric advantages, or to define specific output paths. Used in constraint functions, they can control any unwanted motions of the mechanism. Similarly, the input forces can be used to control the shape of the load-displacement curve, or to limit the amount of force that can be applied. Real mechanical actuators and their guide structures have finite strength, and the optimizer will often attempt to use extremely large input loads if it is allowed to. To prevent this, we place upper-bound constraints on the magnitudes of $F_{in}$ and $F_p$ in each example. The volume fraction is used as an upper-bound constraint to limit the mass of the resulting designs.

In some cases, particularly in those with snap-through behavior or longer structural members loaded in compression, we have encountered optimization convergence issues due to buckling bifurcations in the load-displacement curves. Having multiple stable configurations of a structure is a discontinuity in the design space and can prevent convergence if a different solution is found in each iteration, such as if a member has no strong preference for buckling to one side compared to the other. The same buckling problem is reported by Bruns and Sigmund [52] as a tendency for the optimization to oscillate between structures that do and do not have the buckling response. Li et al. [31] also mention that non-unique solutions are a challenge for considering slender structures with local and global instability. Thus, this issue is not unique to the variable boundary condition formulation of this work. However, the decision to use the guided actuator formulation was motivated by the problem since it adds resistance to buckling by preventing lateral deflection, and this appreciably alleviated the issue for our test cases. In some of the following examples, other strategies to avoid unwanted buckling behaviors have been used as well, such as by placing constraints on the load-displacement curve or by using a larger density filter radius to avoid slender features.

All examples are generated on a desktop computer with unstructured meshes of 3-node triangular finite elements under plane stress conditions. The Method of Moving Asymptotes (MMA) [35] is used as the optimizer with its default settings. Table 1 shows the common parameters used in all of the following examples. For the structure of the compliant mechanisms, we model a flexible rubber-like material by using an elastic modulus of $E_0 = 10$ MPa and a Poisson's ratio of $\nu = 0.49$. A much stiffer material is used for the support structure with $E_s = 2000$ MPa and $\nu_s = 0.3$, giving the shear modulus as $G_s = E_s/(2(1 + \nu_s))$. The SIMP stiffness penalty parameter is set to a constant of $p = 3$. For the strain energy interpolation between linear and nonlinear modeling, we use values for $\beta$ ranging from 500 to 2000 depending on the problem. We use $\rho_0 = 0$ in all problems, as this resulted in better convergence compared with the higher values of this parameter used in the original paper for the method [44]. In the super-Gaussian projection function, we use $b = 2$ so that the radius $r$ corresponds to the contour of 50% density, and a superscript of $P = 4$ to create flat-topped projections with sufficiently smooth edges. The parameter for the smooth minimum and maximum functions is set to $Q = 12$. In the displacement-control algorithm, the convergence tolerance for the residual norm in the corrector loop is set to $1 \times 10^{-6}$ and the maximum number of corrector iterations is set to 20. The topology optimization is considered converged when all constraint functions are satisfied and the average change in the density design variables compared to the previous iteration is less than $1 \times 10^{-4}$ [53].

Table 1 – Optimization Parameters: Common

| Parameter | Symbol | Value |
|---|---|---|
| Design Material Elastic Modulus | $E_0$ | 10 MPa |
| Design Material Minimum Elastic Modulus | $E_{min}$ | $E_0 \times 10^{-9}$ |
| Design Material Poisson's Ratio | $\nu$ | 0.49 |
| Support Material Elastic Modulus | $E_s$ | 2000 MPa |
| Support Material Poisson's Ratio | $\nu_s$ | 0.3 |
| SIMP Penalty | $p$ | 3 |
| Energy Interpolation Threshold | $\rho_0$ | 0 |
| Super-Gaussian Base | $b$ | 2 |
| Super-Gaussian Superscript | $P$ | 4 |
| Smooth Min/Max Superscript | $Q$ | 12 |

In the design plots, the support locations are shown by red dots and the load location is shown by a blue dot. The input displacement direction is shown by the blue arrow. The mesh has been colored red and blue for the supports and load, respectively, and the opacity of the colored meshes represents the relative magnitude of the spring stiffness or load magnitude to visually display how the boundary conditions are applied.

### 6.1. Maximum Displacement

Maximizing the displacement at specified output points given a fixed input force or displacement is a common design objective in topology optimization of compliant mechanisms. A frequently used benchmark problem is a gripper mechanism with a single input force and two supports. In the majority of cases in the literature, a square design domain is used with the load applied at the middle of the left edge and the supports applied at the upper left and lower left corners [4], [33], [34], [46], [54], [55]. We use this configuration for the starting condition, which is illustrated in Figure 5. The design domain is a $10 \times 10$ cm square with a $2 \times 2$ cm square cut away from the center of the right edge to form the jaws of the gripper. The output points are at the upper and lower right corners of the cut-out, and maximizing the output displacement causes these two points to move together as shown by the black arrows. Spring elements of stiffness $k_{out}$ are attached to the output points to simulate the stiffness of a workpiece, which ensures the mechanism can transfer a reasonable amount of load from the input. Fixed non-design regions of solid material are placed along the jaws to guarantee that the entire gripping surface is present in the final design.

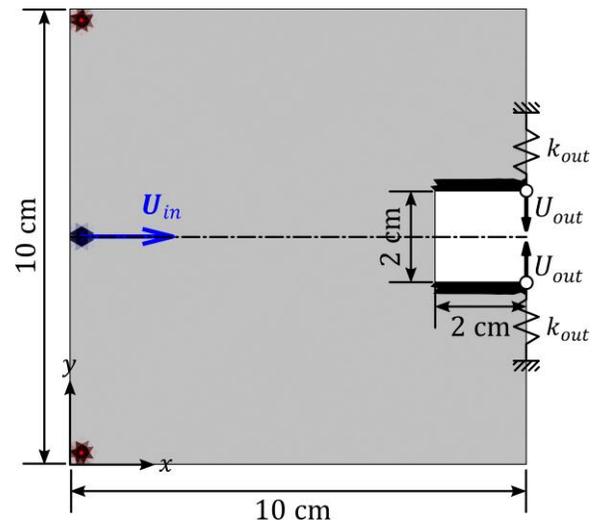

Figure 5 – The initial conditions for the maximum displacement gripper problem.

The optimization parameters specific to this example are listed in Table 2. A mesh of approximately 10,000 elements is used, with the design and support material thicknesses both set to 1 cm. The radius of the density filter and the boundary condition points are each set to 2.5 mm, the actuator stroke is set to 5 mm, and the output springs are given a stiffness of 300 N/m. The move limits, which set the maximum allowed change in the design variables in a single iteration of the optimization, are set to 0.2 for the densities, 2.5 mm for the boundary condition locations, and 5 degrees for the input displacement angle.

Solving for four displacement steps in the nonlinear finite element analysis, the optimization problem consists of maximizing the displacement at the final step. The volume fraction of material is limited to 30%, and the input forces are constrained at each of the four steps such that the upper limit of actuator force increases linearly with input displacement, helping to prevent the optimization from attempting to take advantage of buckling responses. At the maximum stroke length, the maximum allowed input forces are set to 30 N and 7.5 N for the in-line and lateral components, respectively. Mathematically, this optimization problem is given by the statement:

Table 2 – Optimization Parameters: Gripper

| Parameter | Symbol | Value |
|---|---|---|
| Element Size | $h$ | 1.5 mm |
| Number of Elements | $N_e$ | 9,812 |
| Domain Thickness | $t$ | 1 cm |
| Support Thickness | $t_s$ | 1 cm |
| Density Filter Radius | $r_{min}$ | 3 mm |
| Super-Gaussian Radius | $r$ | 2.5 mm |
| Input Displacement Length | $\|\boldsymbol{U}_{in}\|$ | 5 mm |
| Output Spring Stiffness | $k_{out}$ | 300 N/m |
| Energy Interpolation Sharpness | $\beta$ | 500 |
| Move limits: $\boldsymbol{\rho}$ | | 0.2 |
| Move limits: $\boldsymbol{X}_s, \boldsymbol{Y}_s$ | | 2.5 mm |
| Move limits: $X_f, Y_f$ | | 2.5 mm |
| Move limit: $\theta$ | | 5° |

$$\underset{\boldsymbol{\zeta}}{\text{maximize:}} \quad U_{out}^{(4)}$$

$$\text{subject to:} \quad V_f < 0.3,$$

$$F_{in}^{(m)} < \frac{30}{4} m \text{ N}, \qquad m = 1,2,3,4, \qquad (63)$$

$$-\frac{7.5}{4} m \text{ N} < F_p^{(m)} < \frac{7.5}{4} m \text{ N}, \qquad m = 1,2,3,4,$$

where the output displacement objective function has been calculated using two non-zero degrees of freedom in the selector vector of Equation (59), making it a sum of the two output displacements. The minimum and maximum allowed values for the load and support location design variables are set to keep them inside the design domain, at least a distance of $r$ from the edges of the square.

The optimization is performed once with the boundary conditions constrained to remain in their initial conditions, resulting in the design shown in the first row of Figure 6 which achieves an objective function value of $U_{out}^{(4)} = 1.09$ cm. This is a typical outcome for the gripper problem and similar designs can be seen in many other papers [4], [33], [46], [54], [55]. The second row of Figure 6 shows the results after running the optimization with unconstrained boundary conditions. A much different, asymmetrical design is produced where the boundary conditions have moved further inside the domain with the actuator at an angle. For the same input stroke length, the variable boundary condition design achieves an objective function value of $U_{out}^{(4)} = 1.71$ cm, a 57% improvement over the fixed boundary condition design.

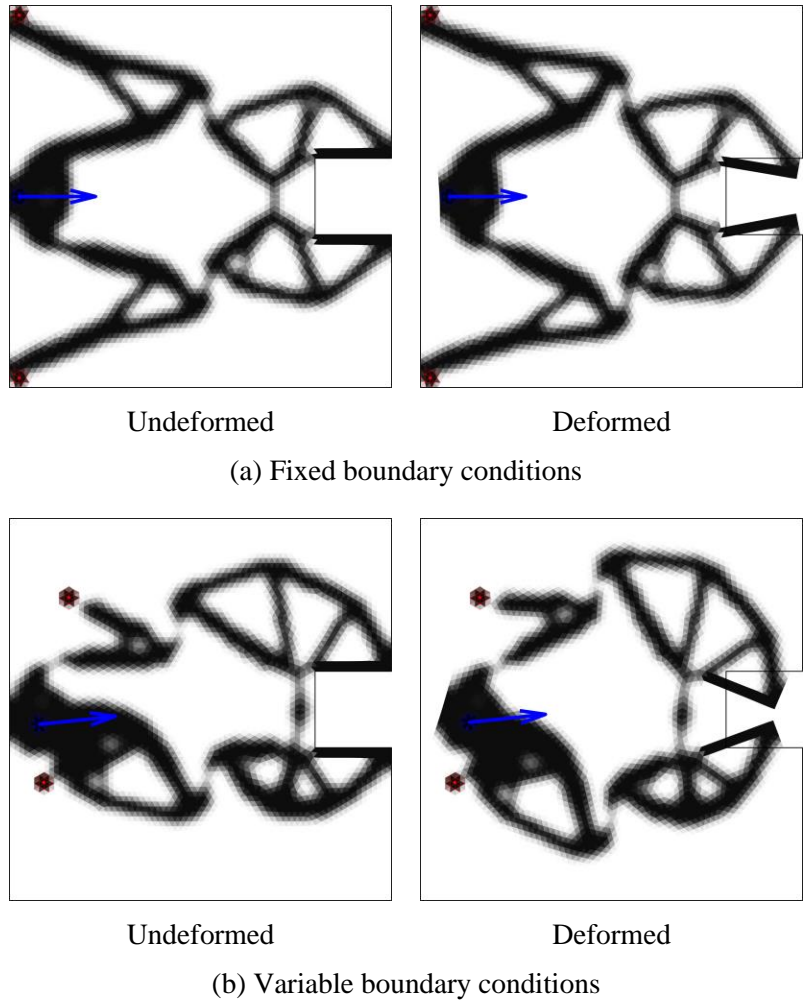

Undeformed        Deformed

(a) Fixed boundary conditions

Undeformed        Deformed

(b) Variable boundary conditions

Figure 6 – Results of the maximum displacement gripper problem. The undeformed and deformed configurations are shown for each design. (a) The design generated with the boundary conditions fixed in their initial configuration. (b) The design generated when allowing the boundary conditions to move from their starting positions.

### 6.2. Bistability

For the bistable structure design problem, we base our test case on the example of a bistable morphing airfoil by Bhattacharyya et al. [30]. The idea of the design problem is to generate a monolithic aileron structure and mechanism that, once actuated, snaps through and passively maintains a high-camber configuration with no additional load input.

Topology optimization of snap-through or bistable structures performs best when using a finite element analysis solution method that is capable of tracing load-displacement curves with snap-back trajectories, such as the arc-length method by Bruns et al. [51]. However, the arc length method as presented in reference [51] is load-controlled and is not able to solve for multiple displacement components, which we need for the guided actuator formulation of this work. Instead, we use our displacement-control algorithm while placing constraints on the load-displacement curve to avoid snap-back behaviors that would otherwise cause the solver to diverge.

The design domain for the problem is a NACA 0012 airfoil with a chord length of 20 cm, shown in Figure 7. The output point is at the trailing edge, with the goal of making it move downwards a distance of 5 mm against the resistance of a spring. The initial boundary conditions are those selected by Bhattacharyya et al.

[30], which represent the best configuration the authors were able to find by manual trial and error. Two supports are placed at the top and bottom edges, at a distance of 6 cm from the leading edge of the airfoil. The actuator is placed on the chord line between the supports and is pointed directly to the right. A third support functions as the aileron hinge and is placed at the bottom edge, 6 cm away from the trailing edge.

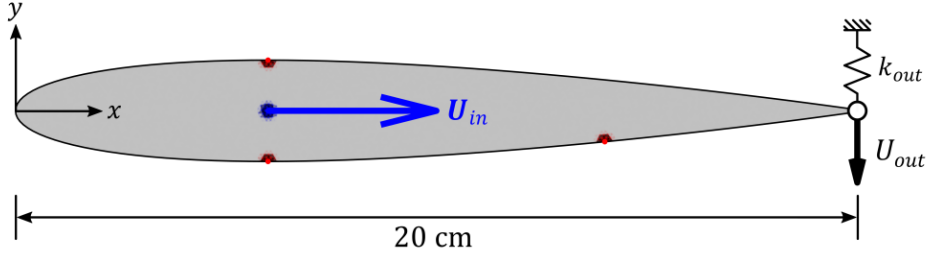

Figure 7 – The initial conditions for the bistable airfoil problem.

The parameters used for the optimization are shown in Table 3. A larger density filter radius is used to avoid slender structures and reduce buckling-related oscillations. The move limits are also reduced to obtain smoother convergence.

Table 3 – Optimization Parameters: Bistable Airfoil

| Parameter | Symbol | Value |
|---|---|---|
| Element Size | $h$ | 1 mm |
| Number of Elements | $N_e$ | 7,680 |
| Domain Thickness | $t$ | 1 cm |
| Support Thickness | $t_s$ | 1 cm |
| Density Filter Radius | $r_{min}$ | 4 mm |
| Super-Gaussian Radius | $r$ | 2 mm |
| Input Displacement Length | $\|U_{in}\|$ | 2.5 mm |
| Output Spring Stiffness | $k_{out}$ | 100 N/m |
| Energy Interpolation Sharpness | $\beta$ | 2000 |
| Move limits: $\boldsymbol{\rho}$ | | 0.05 |
| Move limits: $\boldsymbol{X}_s, \boldsymbol{Y}_s$ | | 0.5 mm |
| Move limits: $X_f, Y_f$ | | 0.5 mm |
| Move limit: $\theta$ | | 1° |

The optimization problem consists of solving for multiple points along the load-displacement curve and minimizing the force of the last point, with the goal of achieving a negative value. Other input forces are constrained to tailor the rest of the force-displacement curve to avoid snap-back, and a constraint is used on the output displacement to cause the deflection of the aileron. Solving for eight evenly distributed displacement points, the formulation we use is written as:

$$
\begin{aligned}
\underset{\boldsymbol{\zeta}}{\text{minimize:}} \quad & F_{in}^{(8)} \\
\text{subject to:} \quad & V_f < 0.4, \\
& U_{out}^{(8)} > 5 \text{ mm}, \\
& F_{in}^{(1)} > 2 \text{ N}, \\
& F_{in}^{(m)} < 15 \sin\left(\pi \frac{m}{6}\right) + 5 \text{ N}, \quad m = 1,2,\dots,6, \\
& -5 \text{ N} < F_p^{(m)} < 5 \text{ N}, \quad m = 1,2,\dots,8.
\end{aligned}
\tag{64}
$$

The upper and lower limits on the load and support locations are set such that they may move outside of the design domain if doing so improves the objective function. This allows the optimizer to completely remove unnecessary supports. The sine wave upper-bound on the input forces $F_{in}^{(m)}$ causes the load-displacement curve to turn downwards earlier than it otherwise would, eliminating a snap-back behavior that caused the displacement-control algorithm to fail. The resulting designs and their convergence plots are shown in Figure 8. Since a snap-through response is itself a buckling behavior, we encountered oscillations in the optimization's convergence. In the study on topology optimization for snap-through mechanisms by Bruns and Sigmund [52], they also report these oscillations and attribute them to the optimization switching between designs that do have a snap-through response and designs that do not.

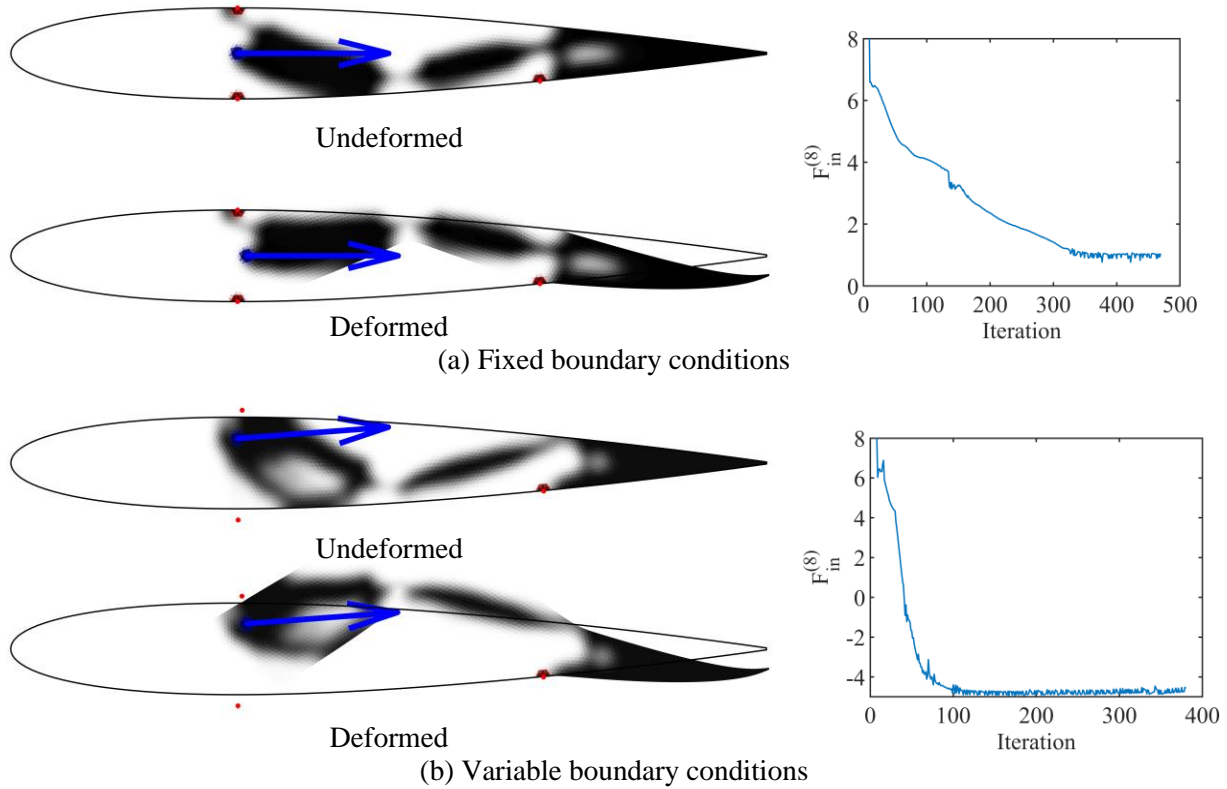

Figure 8 – Results of the bistable airfoil problem. Undeformed configurations, deformed configurations, and optimization convergence plots are shown for each design. (a) The design generated with the boundary conditions fixed in their initial configuration. (b) The design generated when allowing the boundary conditions to move from their starting positions.

The load-displacements curves of each design are shown in Figure 9, which were obtained in a post-analysis where the input displacement was solved in small increments up to a longer stroke length of 4 mm. The curve of the fixed boundary condition design shows that it fails to achieve bistability, with a positive force of 1.0 N at the eighth displacement control point, while the variable boundary condition design successfully reaches a negative force of $-4.6$ N. With the ability to adjust the load and support positions, the optimization increased the geometric advantage of the mechanism by moving the actuator closer to the upper support. With the shorter moment arm around the support, the actuator also must apply a greater force to create the given input displacement, resulting in a stronger snap-through effect. The center point of the upper support also moves slightly outside of the design domain, reducing the size of the supported area and allowing the link to rotate about it more easily.

The results of this problem show that a relatively minor change in boundary conditions can mean the difference between a failed and a successful design. In the previous work of Bhattacharyya et al. [30], the thickness-to-chord ratio of the airfoil had been doubled to give a larger design domain. Here, we used the true NACA 0012 airfoil which is a more difficult optimization problem. Using the configuration that had been successful in the previous designs of [30] did not work, and it was necessary to optimize the boundary conditions to achieve bistability.

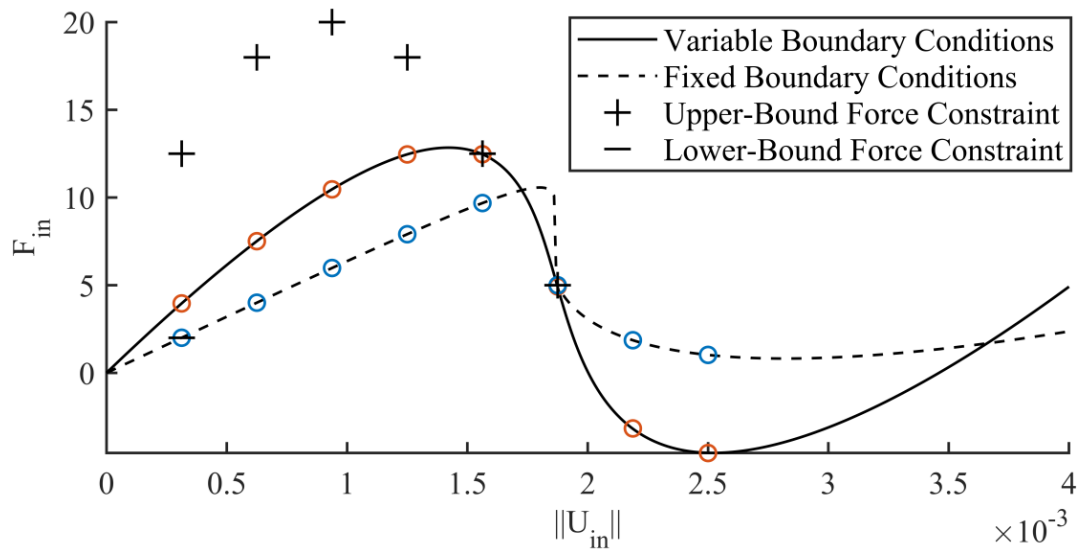

Figure 9 – The force-displacement curves of the bistable airfoil designs. The circular markers show the original eight displacement control points that were solved for in the optimization process.

### 6.3. Path Generation

In path generation problems, the entire path of the output point is specified and the mechanism is optimized to follow that motion as closely as possible. We investigate two test cases based on the horizontal line generation problem from the work of Pederson et al. [33] and the morphing wing design problem from the research of Reinisch et al. [34].

The path generation formulations in these studies work by defining a set of $m$ precision points that the output location of the mechanism should move through at each of the $m$ input displacement states. An error function is used for the objective function which minimizes the difference between the actual path and the desired path. Instead of using springs attached to the output points to achieve structurally stiff mechanisms, the path generation problems use multiple load cases where, in each additional load case $i$, counter forces are applied to the output point to resist its motion. By simultaneously minimizing the output path error in all load cases, fully solid and structurally stiff mechanisms are generated which can perform useful work against external forces.

The design domains and initial conditions for the two test cases are shown in Figure 10. For the horizontal line generator (Figure 10a), a rectangular design domain is used with the output point at the upper-right corner. Four precision points are used to define a straight output path moving 2 cm to the right. Two counter load cases are used, where the first applies a horizontal force $f_1$ pointing to the left and the second applies a vertical force $f_2$ pointing downwards. The boundary conditions are initialized at the bottom edge with the configuration used in [33]. For the morphing wing (Figure 10b), we use the leading 30% of a 20 centimeter long NACA 0012 airfoil as the design domain. The output point and boundary conditions are placed based on the configuration chosen in [34], where two supports represent the wing spar and the guided actuator is placed near the bottom surface of the wing pointing towards the rear. A solid non-design region is placed around the upper part of the domain to create the skin that will bend as the wing changes shape. To keep the skin attached to the spar, a third support is placed slightly above the upper-right corner and kept fixed, even in the variable boundary condition case. A void non-design region is placed between the skin and the interior of the wing to ensure the internal mechanism remains separate from the skin, except at the attachment to the output point. A single precision point is placed 2.5 mm behind and 5 mm below the output point, which represents the position this point on the wing should move to in its morphed state. Two counter load cases are used, with a horizontal force pointing to the right to represent a drag force, and a vertical force pointing upwards representing a lifting force. In each of the two examples, solid non-design regions are placed around the output points to avoid placing the counter forces on low-density elements, which would cause the nonlinear analysis to diverge. The specific parameters for each of these problems are listed in Table 4.

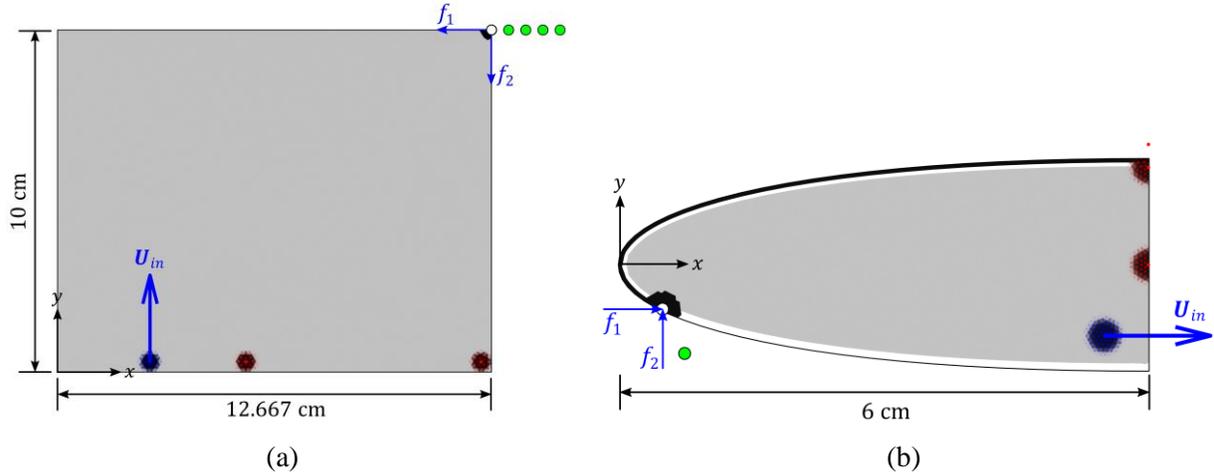

Figure 10 – The initial conditions for the path generation problems. The precision points are shown by the green dots. (a) Horizontal line generator. (b) Morphing wing leading edge.

Table 4 – Optimization Parameters: Path Generation

| Parameter | Symbol | Line Generator | Morphing Wing |
|---|---|---|---|
| Element Size | $h$ | 1.2 mm | 0.5 mm |
| Number of Elements | $N_e$ | 20,290 | 10,828 |
| Domain Thickness | $t$ | 1 cm | 1 cm |
| Support Thickness | $t_s$ | 1 cm | 1 cm |
| Density Filter Radius | $r_{min}$ | 2.4 mm | 1 mm |
| Super-Gaussian Radius | $r$ | 3 mm | 2 mm |
| Energy Interpolation Sharpness | $\beta$ | 500 | 500 |
| Input Displacement Length | $\|\boldsymbol{U}_{in}\|$ | 1 cm | 2 mm |
| Horizontal Counter Force | $f_1$ | 5 N | 1 N |
| Vertical Counter Force | $f_2$ | 5 N | 1 N |
| Move limits: $\boldsymbol{\rho}$ | | 0.2 | 0.2 |
| Move limits: $\boldsymbol{X}_s, \boldsymbol{Y}_s$ | | 3 mm | 1 mm |
| Move limits: $X_f, Y_f$ | | 3 mm | 1 mm |
| Move limit: $\theta$ | | 2° | 5° |

The optimization problem formulation for the horizontal line generator is

$$\underset{\zeta}{\text{minimize:}} \quad \sum_{i=1}^{3} \sum_{m=1}^{4} \left[ \left( X_{out}^{(m,i)} - X_{out}^{*(m)} \right)^2 + \left( Y_{out}^{(m,i)} - Y_{out}^{*(m)} \right)^2 \right]$$

$$\text{subject to:} \quad V_f < 0.2, \tag{65}$$

$$F_{in}^{(m)} < 20 \text{ N}, \qquad\qquad m = 1,2,3,4,$$

$$-5 \text{ N} < F_p^{(m)} < 5 \text{ N}, \qquad\qquad m = 1,2,3,4,$$

where $X_{out}^{(m,i)}, Y_{out}^{(m,i)}$ are the positions of the output point at each input displacement $m$, for each load case $i$, and $X_{out}^{*(m)}, Y_{out}^{*(m)}$ are the corresponding coordinates of the precision points. The load and support locations are also constrained to stay within the design domain, at least a distance $r$ from the edges. Running the problem for the fixed and variable boundary condition cases results in the designs shown in Figure 11. The fixed boundary condition problem generates a design similar to the line generator in the study this problem is based on [33]. For the variable boundary condition design, the actuator and each support move upwards by several centimeters, the actuator and the nearby support move closer to one another horizontally, and the actuator rotates clockwise by 35 degrees. These changes in the boundary conditions lead to a design which more closely follows the straight path of the precision points, giving a value of the path error objective function that is only 43% the value of the fixed boundary condition design's. Using the average distance of the output point from the precision points as a performance metric, the variable boundary condition design's path compared to the fixed boundary condition design's is, on average for all three load cases, only 68% of the distance to the precision points, or 1.47 times closer. To show the improvement visually, the actual output paths of the mechanisms are plotted next to the precision points in Figure 12 by running the finite element analysis on the final designs using small displacement steps. The path of the variable boundary condition design is closer to the straight line of the precision points in the second and third load cases, but is slightly farther for the first load case where the average distance of the output point along the path from the four precision points is 0.55 mm for the fixed boundary condition design and 0.63 mm for the variable boundary condition design. For the second load case, these average distances are 3.1 mm and 1.9 mm, and for the third load case they are 2.2 mm and 1.5 mm.

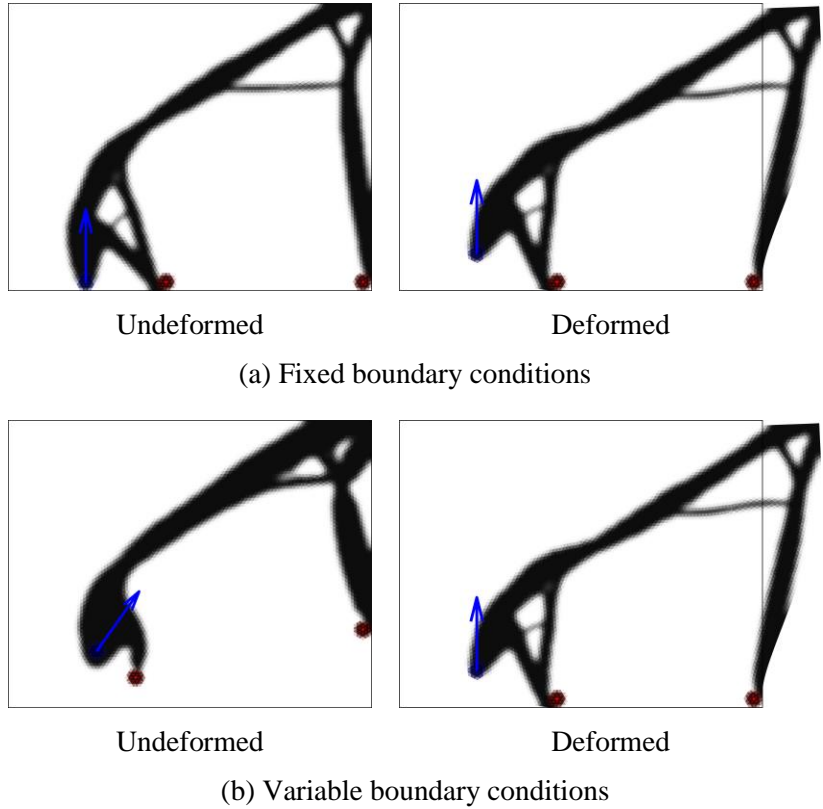

(a) Fixed boundary conditions

(b) Variable boundary conditions

Figure 11 – Results of the horizontal line generator problem. The undeformed and deformed configurations are shown for each design. (a) The design generated with the boundary conditions fixed in their initial configuration. (b) The design generated when allowing the boundary conditions to move from their starting positions.

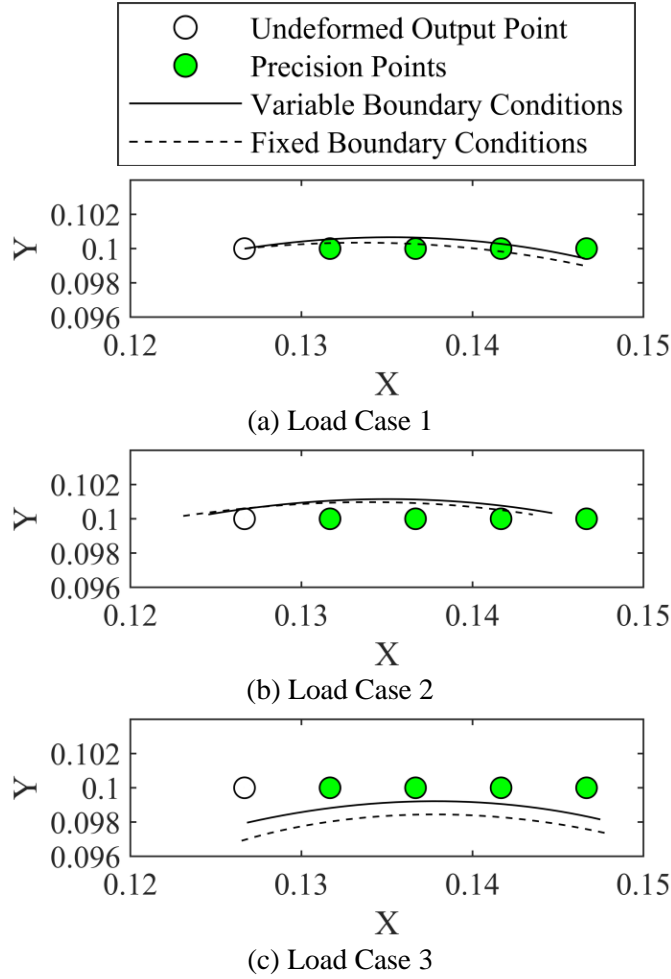

Figure 12 – The actual output paths of the horizontal line generator mechanisms. The units of the axes are in meters. (a) The first load case with no counter forces applied. (b) The second load case with the horizontal counter force $f_1$ applied at the output. (c) The third load case with the vertical counter force $f_2$ applied at the output.

The optimization problem formulation for the morphing wing is written as

$$
\begin{aligned}
\underset{\zeta}{\text{minimize:}} \quad & \sum_{i=1}^{3} \left[ \left( X_{out}^{(i)} - X_{out}^* \right)^2 + \left( Y_{out}^{(i)} - Y_{out}^* \right)^2 \right] \\
\text{subject to:} \quad & V_f < 0.3, \\
& F_{in} < 20 \text{ N}, \\
& -5 \text{ N} < F_p < 5 \text{ N}.
\end{aligned}
\tag{66}
$$

In this problem, the load and support locations are completely unconstrained and may move outside of the design domain if the optimizer finds it advantageous. The results of the optimization for fixed and variable boundary conditions are shown in Figure 13, and the output paths of the optimal designs for each load case are shown in Figure 14. Once again, the fixed boundary condition design is similar to the result of the study the problem is based on [34]. For the variable boundary condition design, the actuator and one of the two supports move closer to the output point, while the other support is moved out of the domain allowing its material to be used in the mechanism's structure instead. The output path passes 4.8 times closer to the precision point, at 21% the distance of the fixed boundary condition design averaged across the three load cases. The paths plotted in Figure 14 show that in the first load case with no counter force the variable boundary condition design morphs to a distance of only 0.11 mm from the target position, while the fixed boundary condition design misses it by eight times that distance, at 0.91 mm from the precision point. In the counter-loaded cases, the variable boundary condition design still comes close to the target position at distances of 0.31 mm for load case two and 0.45 mm for load case three, while the fixed boundary condition design is pushed further away by the counter forces and only comes to 1.4 mm (4.5 times further) and 1.9 mm (4.2 times further) from the precision point in the second and third load cases, respectively.

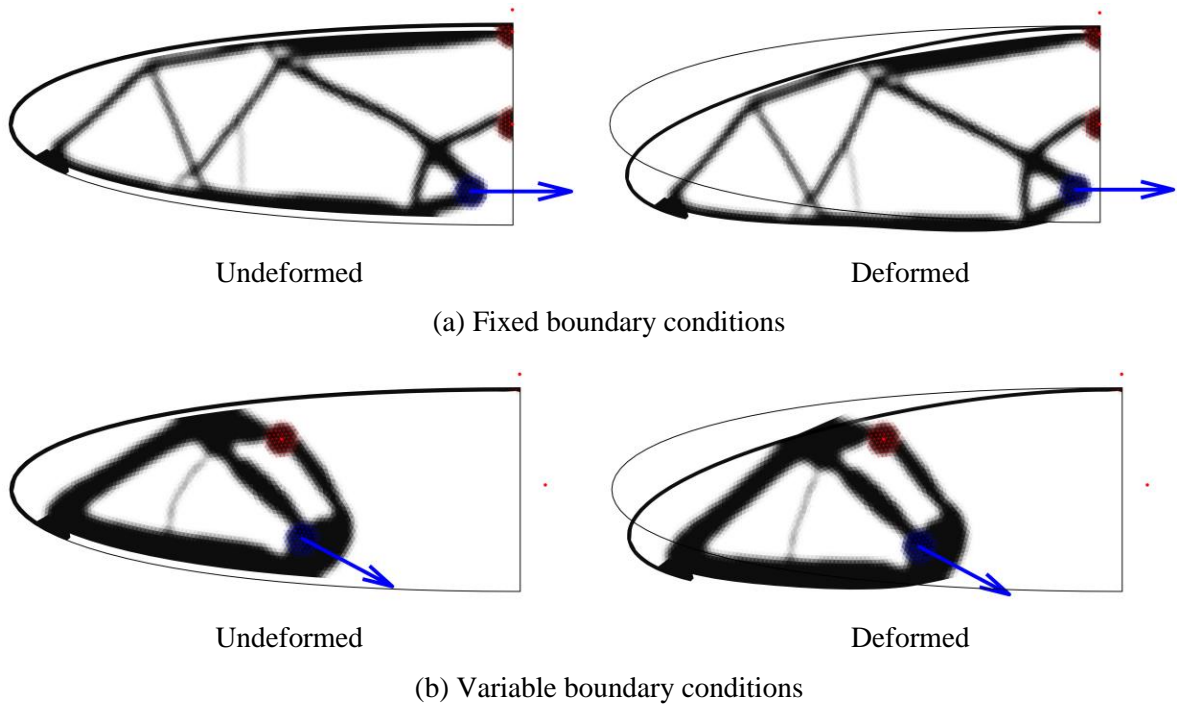

Undeformed          Deformed

(a) Fixed boundary conditions

Undeformed          Deformed

(b) Variable boundary conditions

Figure 13 - Results of the morphing wing leading edge problem. (a) The design generated with the boundary conditions fixed in their initial configuration. (b) The design generated when allowing the boundary conditions to move from their starting positions.

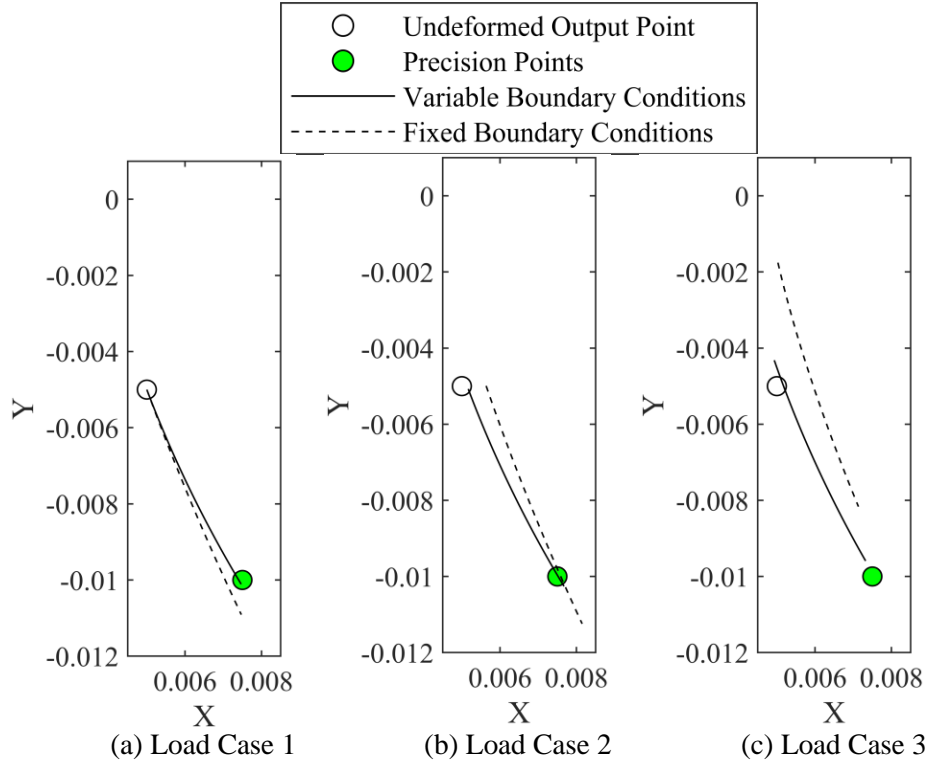

Figure 14 – The actual output paths of the morphing wing leading edge mechanisms. The units of the axes are in meters. (a) The first load case with no counter forces applied. (b) The second load case with the horizontal counter force $f_1$ applied at the output. (c) The third load case with the vertical counter force $f_2$ applied at the output.

For both the line generator and morphing wing variable boundary condition examples, the increases in performance under minimal external loading are due to the better placement of the supports and actuator for achieving the target motions. Comparing the counter-loaded cases to the first load case, the larger increases in distance from the precision points for the fixed boundary condition designs show that the changes in the load and support layouts have also allowed for stiffer optimal structures. These results thus clearly demonstrate the importance of boundary condition placement for both the kinematic and structural aspects of compliant mechanism performance.

## 7. Conclusions

In this work we presented a nonlinear elastic, displacement-controlled topology optimization framework that incorporates the spatial configuration of the boundary conditions as design variables. Expanding on work done for variable loads and supports in linear elasticity [22], we formulated the hyperelastic finite element model, nonlinear governing equations, iterative Newton-Raphson solution algorithm, and adjoint sensitivity equations necessary to achieve a continuously variable input displacement location and orientation. Coupling this capability with variable support locations, we tested the method on several types of nonlinear compliant mechanism synthesis problems. In all examples, using variable boundary conditions produced large improvements over designs obtained using manually-placed fixed boundary conditions. In the bistable morphing airfoil problem, the fixed boundary condition design failed outright to achieve the design goal, while the variable boundary condition design achieved bistability with a relatively small change to the actuator and support positions. These results highlight the difficulty in relying on human intuition for selecting effective boundary conditions and underscore the need for automated methods like the one developed in this work.

Despite the positive results showing the effectiveness of variable boundary conditions, the framework nevertheless has several limitations that should be addressed in future research. The use of a pure displacement-control algorithm prevented us from effectively solving bistable structural optimizations where snap-back behaviors occurred. A different algorithm that is able to trace load-displacement paths through limit points should be developed or extended to be compatible with a variable actuator position. There are also issues associated with nonlinear topology optimization in general. The divergence of unstable low-density elements in the Newton-Raphson iterations is not solved completely by the linear-nonlinear interpolation method [44], which limits the size of displacements and forces that can be applied. We also experienced oscillations in the convergence of optimizations where buckling occurs in the structure. A robust method for avoiding these buckling oscillations would be very useful for nonlinear topology optimization methods in general.

The possibilities for new research directions using variable boundary conditions in topology optimization are numerous. In compliant mechanism design, more complex and counterintuitive mechanisms such as those with multiple degrees of freedom [56], [57] would likely see similar, or potentially greater, benefits than we demonstrated here. Different types of movable mechanical boundary conditions, such as roller supports, have not been attempted yet, which could be applied to design mechanisms with sliding joints. Alternative methods of determining the optimal layouts of the boundary conditions, such as by using free-form design methods instead of the feature-mapping method used here, could be investigated as well. Lastly, optimizing variable boundary conditions for each physics discipline in multiphysics problems holds much potential, for example, in the design of electrothermal actuators [6], [7]. If optimal placements of all six types of boundary conditions including elastic supports, applied forces, applied temperatures, prescribed heat fluxes, applied voltages, and prescribed currents were all simultaneously optimized along with material distribution, it would greatly expand the design space and much better optimal designs may be discovered.

## Acknowledgments

This research was funded by the National Science Foundation through Grant No. 1752045.